\documentclass[aps,prb,twocolumn,showpacs,preprintnumbers]{revtex4-1}
\usepackage{graphicx}
\usepackage{amsmath}
\newcommand{\phon}{\rm phon}
\newcommand{\mg}{\rm mag}
\newcommand{\av}{\mathbf a}
\newcommand{\bv}{\mathbf b}

\begin{document}

\title{Antiferromagnetism of Zn$_2$VO(PO$_4)_2$ and the dilution with Ti$^{4+}$}

\author{A. Yogi}
\author{N. Ahmed}
\affiliation{School of Physics, Indian Institute of Science Education and
Research, Thiruvananthapuram-695016, Kerala, India}
\author{A. A. Tsirlin}
\email{altsirlin@gmail.com}
\affiliation{National Institute of Chemical Physics and Biophysics, 12618 Tallinn, Estonia}
\author{S. Kundu}
\author{A. V. Mahajan}
\affiliation{Department of Physics, Indian Institute of Technology Bombay, Mumbai 400076, India}
\author{J. Sichelschmidt}
\affiliation{Max Planck Institut f\"{u}r Chemische Physik fester
Stoffe, N\"{o}thnitzer Str.~40, 01187 Dresden, Germany}
\author{B. Roy}
\author{Y. Furukawa}
\affiliation{Ames Laboratory and
Department of Physics and Astronomy, Iowa State University, Ames, IA
50011, USA}
\author{R. Nath}
\email{rnath@iisertvm.ac.in}
\affiliation{School of Physics, Indian Institute of Science Education and
Research, Thiruvananthapuram-695016, Kerala, India}

\date{\today }

\begin{abstract}
We report static and dynamic properties of the antiferromagnetic compound Zn$_{2}$(VO)(PO$_{4}$)$_{2}$, and the consequences of non-magnetic Ti$^{4+}$ doping at the V$^{4+}$ site. $^{31}$P nuclear magnetic resonance (NMR) spectra and spin-lattice relaxation rate ($1/T_1$) consistently show the formation of the long-range antiferromagnetic order below $T_N= 3.8-3.9$\,K. The critical exponent
$\beta=0.33 \pm 0.02$ estimated from the temperature dependence of the sublattice magnetization measured by $^{31}$P NMR at 9.4\,MHz is consistent with universality classes of three-dimensional spin models. The isotropic and axial hyperfine couplings between the $^{31}$P nuclei and V$^{4+}$ spins are $A_{\rm hf}^{\rm iso} = (9221 \pm 100)$~Oe/$\mu_{\rm B}$ and $A_{\rm hf}^{\rm ax} = (1010 \pm 50)$~Oe/$\mu_{\rm B}$, respectively. Magnetic susceptibility data above 6.5\,K and heat capacity data above 4.5\,K are well described by quantum Monte-Carlo simulations for the Heisenberg model on the square lattice with $J\simeq 7.7$\,K. This value of $J$ is consistent with the values obtained from the NMR shift, $1/T_1$ and electron spin resonance (ESR) intensity analysis. Doping Zn$_2$VO(PO$_4)_2$ with non-magnetic Ti$^{4+}$ leads to a marginal increase in the $J$ value and the overall dilution of the spin lattice. In contrast to the recent \textit{ab initio} results, we find neither evidence for the monoclinic structural distortion nor signatures of the magnetic one-dimensionality for doped samples with up to 15\% of Ti$^{4+}$. The N\'eel temperature $T_{\rm N}$ decreases linearly with increasing the amount of the non-magnetic dopant.
\end{abstract}


\maketitle

\section{Introduction}
Square lattice of antiferromagnetically coupled Heisenberg spins is the simplest spin model in two dimensions (2D).\cite{chakravarty1989,manousakis1992} Its properties are nowadays well established by extensive numerical studies.\cite{makivic1991,sandvik1997,kim1998} The case of spin-$\frac12$ entails strong quantum effects that reduce the sublattice magnetization\cite{sandvik1997} and have an impact on the correlation length\cite{cuccoli1996,*elstner1995} and spin dynamics.\cite{carretta2000,ronnow2001} The ideal 2D model lacks long-range order (LRO) above zero temperature, following the Mermin-Wagner theorem.\cite{mermin1966} However, any real material features a non-negligible interplane coupling that triggers the LRO at a non-zero temperature $T_N$.\cite{yasuda2005} When interplane couplings are frustrated and inactive, the LRO is driven by anisotropy terms in the spin Hamiltonian.\cite{yildirim1994}

Suppression of the LRO in square-lattice-based magnets is possible via two mechanisms, frustration of in-plane couplings or dilution of the spin lattice. The former mechanism is revealed by the model of the $J_1-J_2$ frustrated square lattice, where the competition between nearest-neighbor couplings $J_1$ and second-neighbor couplings $J_2$ destroys the magnetic order in the vicinity of the quantum critical point at $J_2/J_1=0.5$ for \mbox{spin-$\frac12$}.\cite{[{}][{, and references therein}]schmidt2004} This is the well-established theoretically but hitherto never observed experimentally case of the spin-liquid ground state in 2D.\cite{[{For example: }][{}]darradi2008,*li2012,*jiang2012} The majority of the $J_1-J_2$ systems, mostly layered V$^{4+}$ phosphates, feature columnar antiferromagnetic (AFM) order\cite{bombardi2004,skoulatos2009,nath2009} induced by $J_2>J_1$. Materials with $J_2/J_1<0.5$ developing N\'eel AFM order remain low in number and sometimes challenging to investigate.\cite{tsirlin2008,*oka2008}

The second mechanism is the dilution of the spin lattice with non-magnetic impurity atoms. Diluted systems are largely classical even for spin-$\frac12$.\cite{sandvik2002} The LRO vanishes at the classical percolation threshold of $x_c=0.41$,\cite{kato2000,sandvik2002} where $x$ is the doping level. The doping leads to a gradual suppression of the N\'eel temperature $T_N$,\cite{vajk2002} but in many spin-$\frac12$ materials the $T_N$ drops much faster than expected, because non-magnetic impurity atoms introduce magnetic frustration that contributes to the suppression of the LRO.\cite{liu2009,carretta2011} On the other hand, Li$_2$VOSiO$_4$, which is a frustrated square-lattice antiferromagnet even without dilution,\cite{melzi2000,*melzi2001,rosner2002,*rosner2003,bombardi2004} exhibits a weaker effect on the sublattice magnetization and $T_N$ when diluted with non-magnetic Ti$^{4+}$.\cite{papinutto2005} Apparently, the dilution of real materials never follows the idealized models and entails a modification of individual exchange couplings.

\begin{figure}
\includegraphics{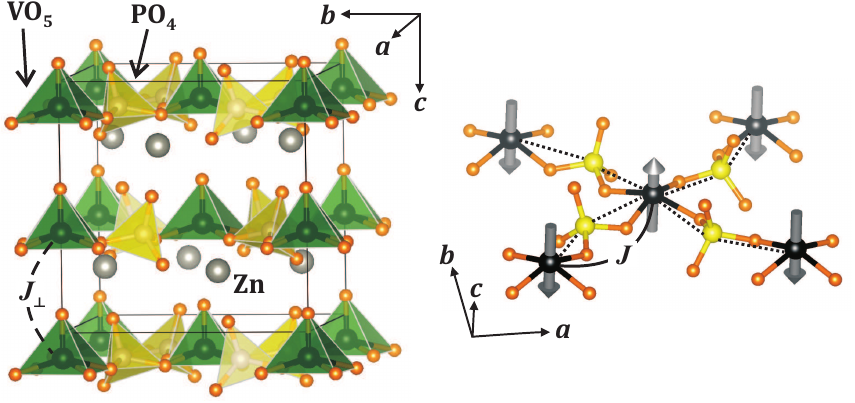}
\caption{\label{fig:structure} Left panel: crystal structure of Zn$_2$VO(PO$_4)_2$. Right panel: magnetic layer in the $ab$ plane. Dotted lines show hyperfine couplings between the P atom and the neighboring V$^{4+}$ spins; each P atom is coupled to two magnetic ions with opposite spin directions. The spins are along the $c$ direction.\cite{yusuf2010}}
\end{figure}

Here, we address magnetic properties and dilution behavior of the spin-$\frac12$ antiferromagnet Zn$_2$VO(PO$_4)_2$, where  magnetic V$^{4+}$ ions can be replaced by the non-magnetic Ti$^{4+}$. The crystal structure of Zn$_2$VO(PO$_4)_2$ features V$^{4+}$O$_5$ pyramids that are linked into layers via PO$_4$ tetrahedra in the $ab$ plane (Fig.~\ref{fig:structure}).\cite{lii1991} Given the small size of the interlayer Zn$^{2+}$ cations, the interplane V--V distance (4.52\,\r A) is shorter than the distance in the $ab$ plane (6.31\,\r A). This led earlier studies\cite{bayi1993} to conclude that Zn$_2$VO(PO$_4)_2$ is a quasi-one-dimensional (1D) magnet with $J_{\perp}\gg J$, where $J$ and $J_{\perp}$ stand for the in-plane and interplane couplings, respectively (Fig.~\ref{fig:structure}). A careful evaluation of thermodynamic data put forward the opposite, quasi-2D scenario with $J\gg J_{\perp}$.\cite{kini2006} Magnetic order observed below $T_N\simeq 3.7$\,K is AFM in the $ab$ plane and ferromagnetic (FM) along the $c$ direction.\cite{yusuf2010} It is consistent with \textit{ab initio} results by Kanungo~\textit{et~al.},\cite{kanungo2013} who also addressed the diluted, Ti-doped case and proposed that the 25\% Ti doping should induce a monoclinic distortion reinstating the 1D physics, but this time in the $ab$ plane and not along the $c$ direction.

In this study, we attempt to verify the prediction\cite{kanungo2013} concerning the Ti-doped Zn$_2$VO(PO$_4)_2$ experimentally, and show that within the feasible doping levels, neither the monoclinic distortion nor the 1D physics are observed. Instead, thermodynamics of Ti-doped Zn$_2$VO(PO$_4)_2$ above $T_N$ is largely consistent with expectations for the diluted square lattice of Heisenberg spins. The N\'eel temperature is systematically suppressed upon the dilution, and the rate of suppression is similar to that in Li$_2$VOSiO$_4$. We provide accurate estimates of the in-plane exchange coupling in order to assess this effect quantitatively, and discuss our results in the light of available experimental data on diluted AFM square lattices. We also report additional characterization for thermodynamic properties, ground state and spin dynamics of the parent Zn$_2$VO(PO$_4)_2$ compound. These data will serve as a starting point for detailed studies of the doped material.

\section{Methods}
\label{sec:methods}
Polycrystalline samples of Zn$_2$(V$_{1-x}$Ti$_x$)O(PO$_4$)$_2$ ($x$ = 0\%, 5\%, 10\%, and 15\%) were prepared by a conventional solid-state reaction route. In the first step, Zn$_2$P$_2$O$_7$ was synthesized using ZnO (Aldrich, 99.999\%) and NH$_4$H$_2$PO$_4$ (Aldrich, 99.999\%) as starting materials. The stoichiometric mixtures were fired at $600$\,$^{\circ}$C in air with one intermediate grinding. In the second step, Zn$_2$P$_2$O$_7$ was mixed with VO$_2$ (Aldrich, 99.999\%) and TiO$_2$ (Aldrich, 99.999\%) and then the stoichiometric mixtures were fired in flowing Ar-gas atmosphere with several intermediate grindings and palletizations at $850$\,$^{\circ}$C.

To check the sample purity, powder x-ray diffraction (XRD, PANalytical powder diffractometer and CuK$_{\alpha}$ radiation, $\lambda_{\rm ave}=1.54182$\,\AA) was performed at room temperature. The samples with $x$ = 0\%, 5\%, and 10\%, were single-phase, but at higher doping concentrations several impurity phases including Ti$_4$O$_3$(PO$_4$)$_3$ emerged. Our repeated attempts to achieve higher doping levels by increasing or lowering the firing temperature were unsuccessful. Therefore, we focus on studying the samples with $x\leq 15$\,\%, where a minor amount of non-magnetic Ti-containing impurities does not hinder the data analysis.

Le Bail fit of the powder XRD data was performed using the \texttt{FullProf} software package based on the tetragonal structure
with space group $I4cm$ to determine the lattice parameters.\cite{fullprof} No indications of a symmetry lowering were observed. All the data sets could be fitted using structural data of the parent compound as the initial parameters. The refined lattice parameters and the goodness of fits ($\chi^2$) are listed in Table~\ref{tab:latticeparameters}. No significant change in lattice constants ($a$ and $c$) and unit cell volume ($V$) was observed with increasing $x$. Given the fact that Ti$^{4+}$ features nearly the same ionic radius (0.51~\AA) as V$^{4+}$ (0.53~\AA), we do not expect any substantial changes in the cell volume. Thus our experimental observation is consistent with expectations based on the ionic radii.
\begin{table}
\caption{\label{tab:latticeparameters}
Lattice parameters ($a$, $c$, and $V$) and the goodness of fit ($\chi^2$) obtained from the Le Bail fit of the powder XRD data for Zn$_2$V$_{1-x}$Ti$_x$O(PO$_4)_2$.}
\begin{ruledtabular}
\begin{tabular}{c@{\hspace{1em}}ccc@{\hspace{1em}}c}
  $x$ & $a$~(\r A) &  $c$~(\r A)  &      $V$~(\r A)$^{3}$  & $\chi^2$~(\%)  \\\hline
 0.00 & 8.9221(2)       &  9.0376(2)         &      719.44(2)               &     1.83       \\
 0.05 & 8.9218(2)       &  9.0326(2)         &      718.99(2)               &     2.11       \\
 0.10 & 8.9243(3)       &  9.0287(3)         &      719.07(4)               &     4.75       \\
 0.15 & 8.9251(3)       &  9.0292(3)         &      719.24(4)               &     5.49       \\
\end{tabular}
\end{ruledtabular}
\end{table}

Temperature ($T$) dependent magnetic susceptibility $\chi(T)$ and heat capacity $C_p(T)$ measurements were performed using a commercial Physical Property Measurement System (PPMS, Quantum Design). For the $\chi(T)$ measurement, the vibrating sample magnetometer (VSM) attachment to the PPMS was used. $C_p (T)$ was measured by the relaxation technique on a pressed pellet using the heat capacity option of the PPMS.

Electron spin resonance (ESR) experiments were carried out on a fine-powdered sample with a standard continuous-wave spectrometer between 5\,K and 300\,K. We measured the power $P$ absorbed by the sample from a transverse magnetic microwave field (X-band, $\nu\simeq 9.4$\,GHz) as a function of an external, static magnetic field $H$. A lock-in technique was used to improve the signal-to-noise ratio which yields the derivative of the resonance signal $dP/dB$.

The NMR measurements were carried out using pulsed NMR techniques on $^{31}$P (nuclear spin $I=1/2$ and gyromagnetic ratio $\gamma_{N}/2\pi = 17.237$\,MHz/T) nuclei in the temperature range 1.5\,K $\leq T \leq $ 250\,K. We have carried out the NMR measurements at two different radio frequencies of $75.5$\,MHz and $9.4$\,MHz that correspond to applied fields of about $4.38$\,T and $0.545$\,T, respectively. Spectra were obtained either by Fourier transform of the NMR echo signals or by sweeping the field at a fixed frequency. The NMR shift $K(T)=(H_{\rm ref}-H(T))/H(T)$ was determined by measuring the resonance field of the sample [$H(T)$] with respect to nonmagnetic reference H$_{3}$PO$_{4}$ (resonance field $H_{\rm ref}$). The $^{31}$P spin-lattice relaxation rate $1/T_{1}$ was measured by the conventional single saturation pulse method.

Magnetic susceptibility and specific heat of the pristine and diluted AFM square lattice of Heisenberg spins was obtained from quantum Monte-Carlo (QMC) simulations performed by the \texttt{loop} algorithm\cite{loop} of the \texttt{ALPS} simulation package.\cite{alps} Simulations were performed on $L\times L$ finite lattices with periodic boundary conditions and $L$ up to 80. For the three-dimensional (3D) model of coupled square planes, the $16\times 16\times 8$ finite lattice was used. Finite-size effects are negligible in the temperature range considered ($T/J\geq 0.6$).

\section{Pure Zn$_2$VO(PO$_4$)$_2$}

\subsection{Thermodynamic properties}
\label{sec:thermo}
\begin{figure}
\includegraphics{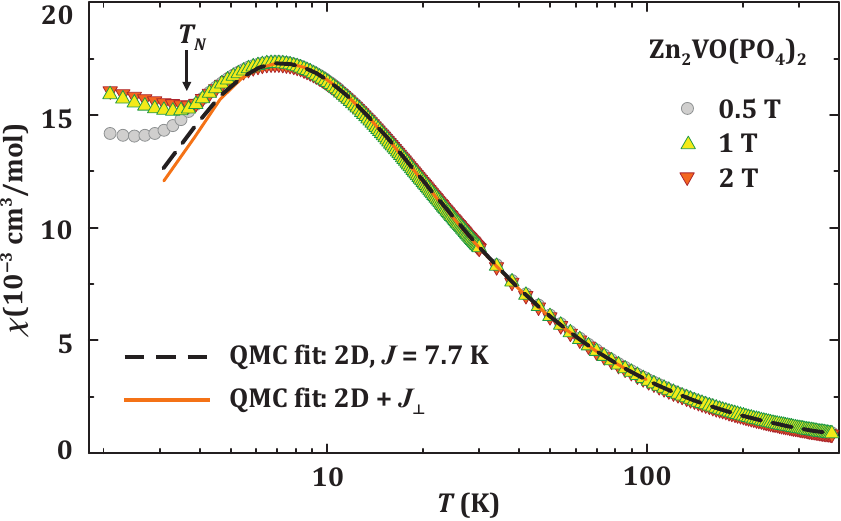}
\caption{\label{fig:chi} (Color online) Magnetic susceptibility ($\chi$) of Zn$_2$VO(PO$_4$)$_2$ measured in the applied fields  $\mu_0H=0.5$\,T, 1\,T, and 2\,T. The dashed line is the QMC fit of the 1\,T data with the 2D model. The solid line is the QMC fit with the 3D model featuring the interplane coupling $J_{\perp}/J=-0.1$ (see text for details). The arrow marks the N\'eel temperature $T_N$, where the data measured at 0.5\,T and at higher fields diverge because of the spin-flop transition.}
\end{figure}
In order to analyze the effect of doping on the exchange couplings, we first consider thermodynamic properties of the parent compound. Our $\chi$ and $C_p$ data are similar to those reported by Kini~\textit{et~al.}\cite{kini2006} Magnetic susceptibility (Fig.~\ref{fig:chi}) shows a broad maximum around 6.9\,K corresponding to the short-range order in 2D. The LRO transition manifests itself by a kink at $T_N\simeq 3.8$\,K in the susceptibility data measured at 1\,T and 2\,T. This effect is due to the spin-flop transition that increases the susceptibility below $T_N$.

\begin{figure}
\includegraphics{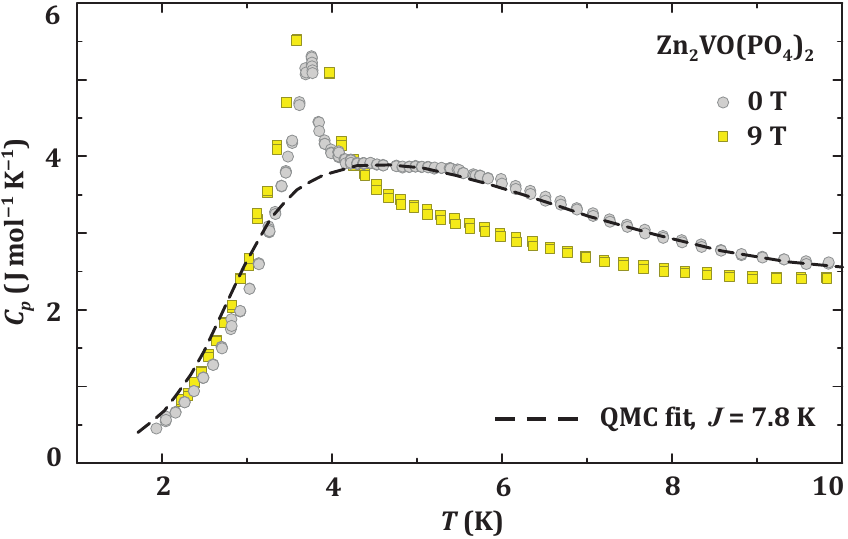}
\caption{\label{fig:heat} (Color online) Specific heat ($C_p$) of Zn$_2$VO(PO$_4$)$_2$ measured in the applied fields  $\mu_0H=0$\,T and 9\,T. The dashed line is the QMC fit of the zero-field data. Magnetic field shifts the entropy from the broad maximum above $T_N$ to the transition anomaly at $T_N$.}
\end{figure}

Specific heat reveals a $\lambda$-type anomaly at $T_N$ (Fig.~\ref{fig:heat}). The hump above $T_N$ is a signature of the broad maximum related to the 2D short-range order. The $C_p(T)$ flattens out around 10\,K and increases at higher temperatures because of the growing phonon contribution. Applied magnetic field suppresses the hump and shifts the entropy to the transition anomaly at $T_N$. However, the value of $T_N=3.8-3.9$\,K remains unchanged.

Magnetic susceptibility of Zn$_2$VO(PO$_4)_2$ is well described by the AFM square-lattice model. The susceptibility simulated by QMC was fitted to the experimental curve using the expression:
\begin{equation}
  \chi=\chi^*\times\left(\frac{N_Ak_Bg^2}{J}\right),
\end{equation}
where $\chi^*$ is the reduced susceptibility calculated by QMC, $N_A$ is Avogadro's number, $k_B$ is Boltzmann constant, and $g$ is the $g$-factor. We fit the data with $J=7.7$\,K and $g=1.95$ down to 6.5\,K (Fig.~\ref{fig:chi}). At lower temperatures, experimental susceptibility lies above the simulated curve. This deviation can be mitigated by decreasing the $J$ value to 7.4\,K. However, the description of the high-temperature part is deteriorated, and the $g$-value drops to 1.91, which is below our ESR estimate (Sec.~\ref{sec:esr}) and below the typical range of powder-averaged $\bar g=1.94-1.98$ reported for V$^{4+}$ oxide compounds.\cite{tsirlin2011,forster2013,forster2014} A Curie-like impurity contribution also improves the fit in the low-temperature region, but introduces discrepancies at higher temperatures. Moreover, the low-field data measured at 0.5\,T do not show any signatures of a Curie-like upturn down to 2\,K (Fig.~\ref{fig:chi}).

Specific heat above $T_N$ is also consistent with the predictions of the square-lattice model. For a proper comparison magnetic ($C_{\mg}$) and phonon ($C_{\phon}$) contributions to the specific heat should be separated. Unfortunately, a non-magnetic reference compound for our case is not available, because not more than 15\% of Ti$^{4+}$ can be doped into Zn$_2$VO(PO$_4)_2$, and the hypothetic end member Zn$_2$TiO(PO$_4)_2$ does not exist. Kini~\textit{et~al}.\cite{kini2006} approximated $C_{\phon}$ with a series of Debye functions and demonstrated that $C_{\phon}<C_{\mg}$ below 10\,K. By using the data from Ref.~\onlinecite{kini2006}, we verified that in this temperature range of our interest $C_{\phon}$ follows the $T^3$ behavior. Therefore, we fitted our data as:
\begin{equation}
  C_p^{\rm exp}=C_p^{\rm QMC}R+\beta T^3,
\end{equation}
where $R$ is the gas constant, and $\beta$ is treated as an adjustable parameter, because in doped samples it may change following the change in the atomic masses and the formation of defects having influence on phonons. This way, we compare specific heat of Zn$_2$VO(PO$_4)_2$ to the QMC result and find best agreement for $J=7.8$\,K (Fig.~\ref{fig:heat}) that is nearly equal to $J=7.7$\,K from the susceptibility fit.

For the sake of completeness, let us discuss possible deviations from the idealized square-lattice model. The ratio $T_N/J\simeq 0.51$ implies $|J_{\perp}|/J\simeq 0.1$.\cite{yasuda2005} Although the N\'eel temperature of Zn$_2$VO(PO$_4)_2$ is rather high for a quasi-2D magnet, strong signatures of the 2D physics have been observed experimentally at $T>T_N$. Apart from the excellent description of both magnetic susceptibility (Fig.~\ref{fig:chi}) and specific heat (Fig.~\ref{fig:heat}) with the purely 2D models, neutron studies revealed Warren-type diffuse scattering above $T_N$, which is indicative of 2D spin correlations.\cite{yusuf2010} Therefore, Zn$_2$VO(PO$_4)_2$ can be classified as an intermediate case between the quasi-2D and spatially anisotropic 3D magnets. However, even with the realistic interlayer coupling ($|J_{\perp}|/J\simeq 0.1$) included in the model, no improvement of the susceptibility fit could be achieved. The susceptibility of the 3D model deviates from that of the 2D model only below 5\,K when the magnetic ordering transition at $T_N$ is approached (Fig.~\ref{fig:chi}).

The in-plane square lattice in Zn$_2$VO(PO$_4)_2$ can be weakly frustrated by the second-neighbor coupling $J_2$. Frustrated spin models are not amenable to QMC simulations because of the notorious sign problem. Therefore, we resort to the high-temperature series expansion (HTSE) of the frustrated square lattice model\cite{rosner2002,*rosner2003} for the magnetic susceptibility that is generally valid at temperatures exceeding individual magnetic couplings $J_i$. The data above 10\,K yield $J\simeq 7.8$\,K, $J_2\simeq 0.3$\,K, and $g\simeq 1.96$ in excellent agreement with Ref.~\onlinecite{kini2006}. Therefore, the frustration of the square lattice in Zn$_2$VO(PO$_4)_2$ is extremely weak, $J_2/J\simeq 0.04$ to be compared with the \textit{ab initio} result $J_2/J=(t_2/t_1)^2\simeq 0.03$ from Ref.~\onlinecite{kanungo2013}. We do not expect that this weak frustration affects thermodynamic properties.

The remaining source of the marginal discrepancy between the square-lattice model and the experimental magnetic susceptibility is the magnetic anisotropy. However, we do not find any strong signatures of the anisotropy in the NMR data reported below.

\subsection{ESR}
\label{sec:esr}
\begin{figure}
\includegraphics [width=3.5in]{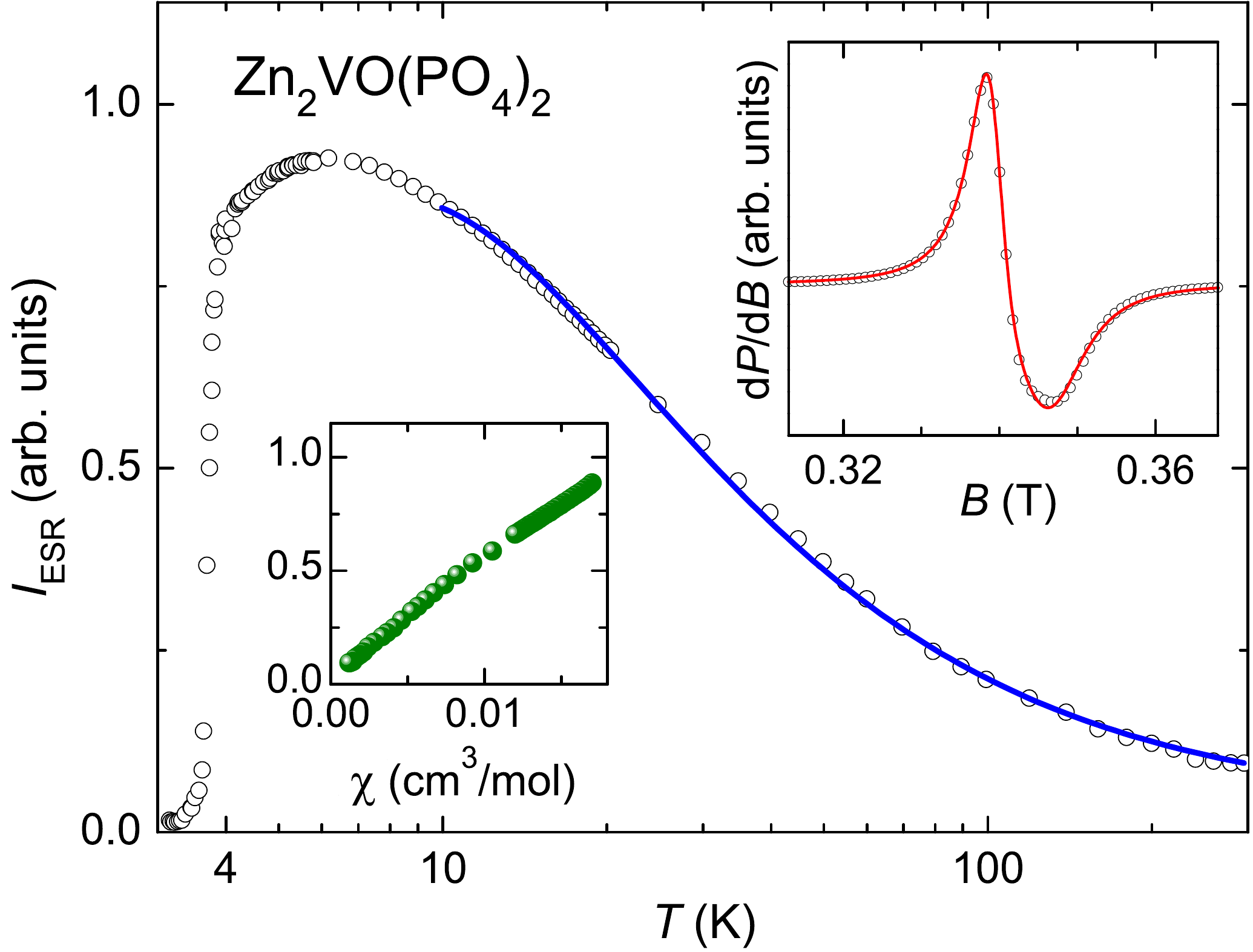}
\caption{\label{esr} (Color online) Temperature dependent ESR intensity, $I_{\rm ESR}(T)$, obtained by double integration of the ESR spectra of powdered Zn$_2$VO(PO$_4$)$_2$ sample; solid line represents the fit described in the text. Upper right inset shows a typical spectrum (symbols) together with a Lorentzian shape (solid line) powder-averaged for a uniaxial $g$-factor anisotropy. Bottom left inset shows the relation between $I_{\rm ESR}(T)$ and $\chi$ measured at a field of 0.5~T and temperatures between 9~K and 295~K.}
\end{figure}
Results of the ESR experiment are presented in Fig.~\ref{esr}. In the right inset of Fig.~\ref{esr}, a typical ESR spectrum at room temperature is shown. The shape of the spectra can be well described by a powder-averaged Lorentzian line for the case of an easy-axis anisotropy of the $g$-tensor, as shown by the solid line, yielding the parallel $g_{\parallel}=1.94(6)$ and perpendicular $g_{\perp}=1.98(7)$ components at $T=295$\,K. The isotropic
$g=\sqrt{(g^{2}_{\parallel}+2g^{2}_{\perp})/3}$ was calculated to be $\sim1.97$. Regardless of $g_{\parallel} < g_{\perp}$ (as expected for an easy-axis anisotropy), these V$^{4+}$ $g$-factors are similar to those reported for Pb$_2$VO(PO$_4$)$_2$ (Ref.~\onlinecite{forster2013}) or SrZnVO(PO$_4$)$_2$ (Ref.~\onlinecite{forster2014}).

The integrated ESR intensity [$I_{\rm ESR}(T)$] increases with decreasing temperature and then exhibits a broad maximum at about 7\,K as observed in $\chi(T)$ (Fig.~\ref{fig:chi}) and $K(T)$ (Fig.~\ref{K}). Below $T_{\rm N}$, it decreases rapidly towards zero. $I_{\rm ESR}(T)$, as obtained by integrating the whole spectrum, linearly depends on the uniform static susceptibility $\chi(T)$ of the V$^{4+}$ spins probed by ESR. Hence, one can get an estimate of the exchange couplings by fitting $I_{\rm ESR}(T)$ data to the HTSE of the square lattice model. We fitted the data above 8\,K to $I_{\rm ESR}(T) = A+B\times \chi_{\rm spin}$, where $A$ and $B$ are arbitrary constants and $\chi_{\rm spin}$ is the expression for HTSE (valid over $\frac{k_B T}{J} \gtrsim 0.7$) of $\chi(T)$ for
the 2D $S = 1/2$ HAF square lattice given by Rushbrooke and Wood\cite{rushbrooke1958}
which can be written as
\begin{eqnarray}
\chi_{\rm spin}(T) &=& \frac{N_A\mu_B^2g^2}{J} \nonumber\\
&\times& [(4x+4+2.00025x^{-1}+0.66656x^{-2}+0.06286x^{-3} \nonumber\\
&-& 0.060434x^{-4}+0.000237x^{-5}]^{-1},
\label{2D}
\end{eqnarray}
where $x=\frac{k_{B}T}{J}$.
By fitting the data in the high-$T$ regime ($T > 8$~K), the exchange coupling was estimated to be $J=(8.7 \pm 0.2)$\,K which agrees with the values estimated from $\chi$ and NMR shift (discussed later) analysis.
In an attempt to see how $I_{\rm ESR}$ scales with $\chi$, we plotted $I_{\rm ESR}$ vs. $\chi$ with temperature as an implicit parameter (see bottom left inset of Fig.~\ref{esr}). A nearly linear behavior down to 9\,K reflects that $I_{\rm ESR}(T)$ tracks $\chi(T)$ of the V$^{4+}$ spins very well.

The influence of critical spin fluctuations on the temperature dependencies of linewidth and resonance field become noticeable below 30~K. However, we refrained from using these temperature dependencies to obtain information on the critical spin dynamics for which an accurate determination of the parallel and perpendicular line components is needed. For this purpose, our powder spectra are too broad compared to the difference between $g_{\parallel}$ and $g_{\perp}$. Investigations of single crystals would certainly provide the required accuracy as the ESR results on Pb$_2$VO(PO$_4$)$_2$ have shown in Ref.~\onlinecite{forster2013}.

\subsection{$^{31}$P NMR Shift}
\begin{figure}
\includegraphics [width=3in]{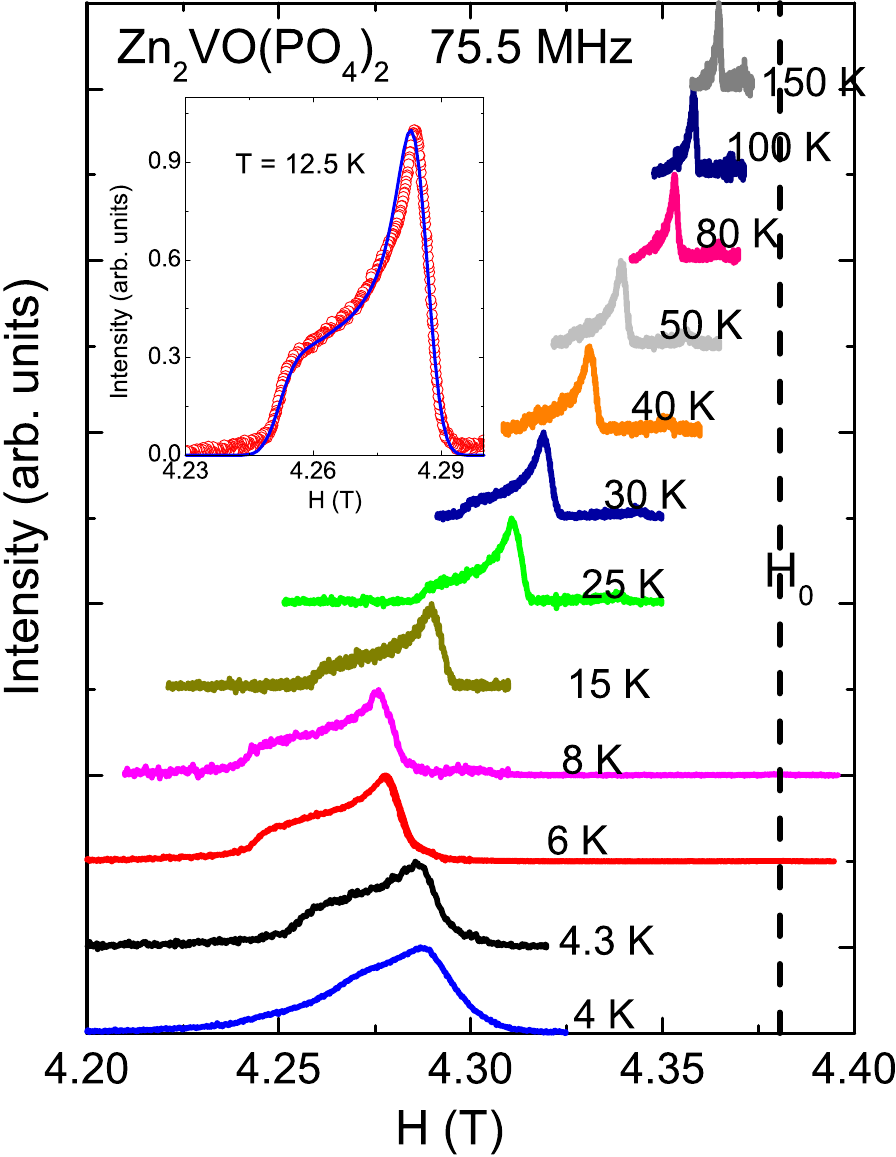}
\caption{\label{spk} (Color online) Field-sweep $^{31}$P NMR spectra at different temperatures $T$ ($T > T_{\rm N}$) for polycrystalline Zn$_{2}$(VO)(PO$_{4}$)$_{2}$ measured at 75.5\,MHz. The vertical dashed line corresponds to the $^{31}$P resonance frequency of the reference sample H$_{3}$PO$_{4}$. Inset shows the $^{31}$P NMR spectrum at $12.5$\,K (open circles). The solid line is the fit. The NMR shift values obtained from the fitting are $K_{\rm iso} \simeq 2.47\%$ and $K_{\rm ax} \simeq 0.27\%$.}
\end{figure}

\begin{figure}
\includegraphics [width=4.5in]{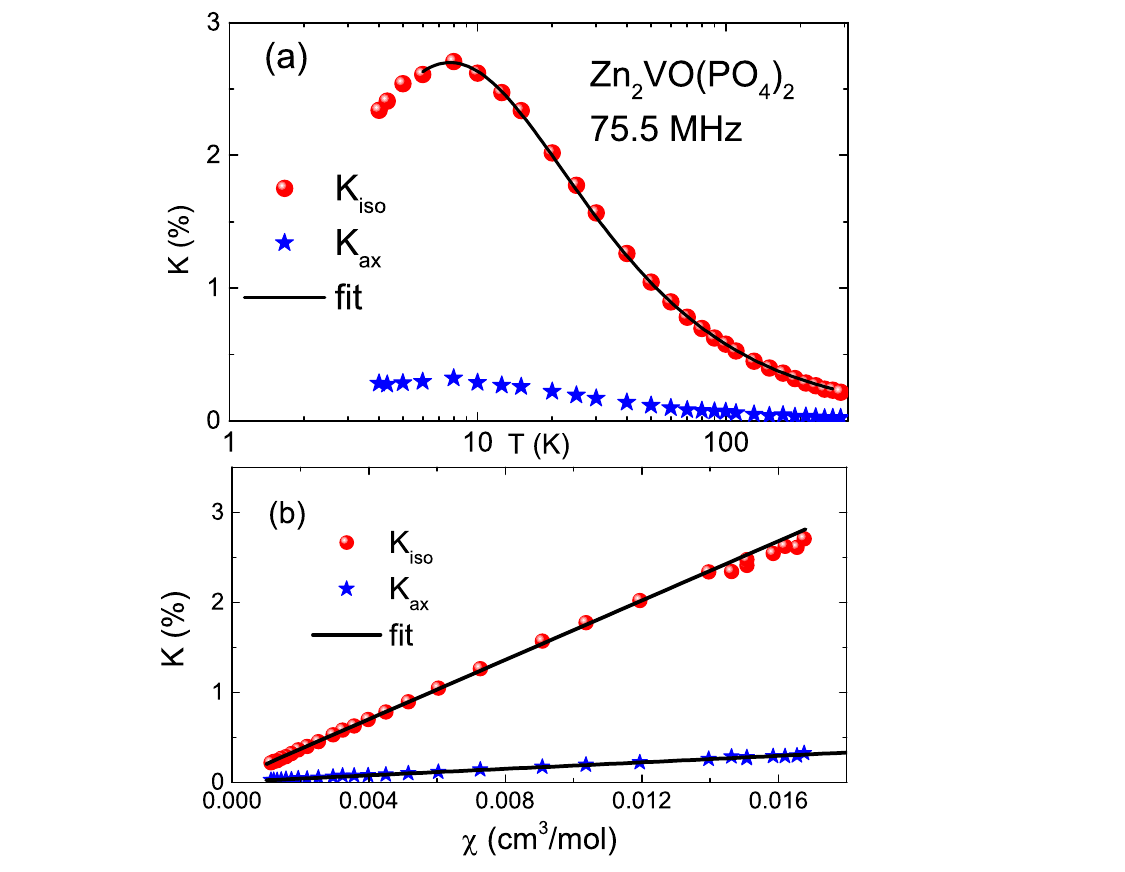}
\caption{\label{K} (Color online) (a) Temperature-dependent NMR shift $K$ vs. $T$. The solid line is the fit of $K_{\rm iso}$ by Eq.~\eqref{shift}. (b) $^{31}$P shift $K$ vs. $\chi$ measured at 2\,T is plotted with temperature as an implicit parameter for both $K_{\rm iso}$ and $K_{\rm ax}$. The solid lines are linear fits.}
\end{figure}

According to Ref.~\onlinecite{lii1991}, the structure of Zn$_{2}$VO(PO$_{4}$)$_{2}$ features one P site. We observed a narrow spectral line above $T_{\rm N}$ as is expected for an $I=1/2$ nucleus.\cite{nath2005,*nath2008,*nath2008b} Figure~\ref{spk} shows the $^{31}$P NMR spectra measured at different temperatures. The line shape was found to be asymmetric because of the anisotropy in $\chi(T)$ and/or in the hyperfine coupling constant between the P nucleus and the V$^{4+}$ spins.

The line position was found to shift with temperature. Temperature dependence of the NMR shift $K$ extracted by fitting the spectra (see inset of Fig.~\ref{spk}) are presented in Fig.~\ref{K}(a), which shows a strong anisotropy along different directions. At high temperatures, both isotropic ($K_{\rm iso}$) and axial ($K_{\rm ax}$) parts of the NMR shift vary in
a Curie-Weiss manner and then pass through a broad maximum at around 9\,K reflecting the 2D short-range order, similar to the $\chi(T)$ data (Fig.~\ref{fig:chi}).

The NMR shift $K(T)$ is related to the spin susceptibility $\chi_{\rm spin}(T)$ by the relation
\begin{equation}
K(T)=K_{0}+\frac{A_{\rm hf}}{N_{\rm A}} \chi_{\rm spin}(T),
\label{shift}
\end{equation}
where $K_{0}$ is the temperature-independent chemical shift, and $A_{\rm hf}$ is the hyperfine coupling constant between the P nuclei and the V$^{4+}$ electronic spins. The $K$ vs.\ $\chi$ plot with $T$ as an implicit parameter is fitted very
well by a straight line [Fig.~\ref{K}(b)] over the whole temperature range ($T > T_{\rm N}$) yielding the isotropic and axial parts of the hyperfine coupling $A_{\rm hf}^{\rm iso} = (9221 \pm 100)$ and $A_{\rm hf}^{\rm ax} = (1010 \pm 50)$~Oe/$\mu_{\rm B}$, respectively. Since the NMR shift is a direct measure of $\chi _{\rm spin}$ and is free from extrinsic impurities, it serves as an independent test for the bulk susceptibility $\chi(T)$. We fitted the temperature dependence of $K_{\rm iso}$ above 6~K by Eq.~\eqref{shift} where the expression for $\chi_{\rm spin}$ is given in Eq.~\eqref{2D}. During the fitting process, $g$ and $A_{\rm hf}^{\rm iso}$ were fixed to the values $g \simeq 1.97$ and $A_{\rm hf}^{\rm iso} \simeq 9221$~Oe/$\mu_{\rm B}$, obtained from the ESR experiments and $K_{\rm iso}$ vs. $\chi$ analysis, respectively.
In this way, we obtained $K_{0} = (0.025 \pm 0.001)$~\% and $J/k_{\rm B} = (8.4 \pm 0.3)$~K. The fit is shown in
Fig.~\ref{K}(a) as a solid line. The resulting $J$ value is close to the values
estimated from the $\chi(T)$ analysis\cite{kini2006} and neutron diffraction experiments.\cite{yusuf2010}

\begin{figure}
\includegraphics [width=3in]{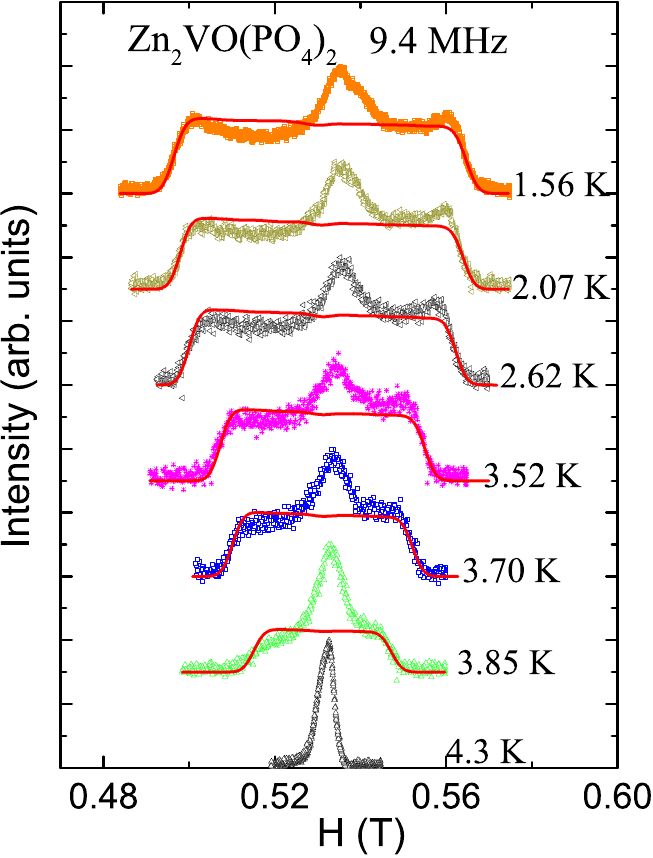}
\caption{\label{belowTn} (Color online) Temperature-dependent $^{31}$P NMR spectra measured at 9.4\,MHz. The solid lines are the fits to the spectra at different temperatures as in Ref.~\onlinecite{nath2014}. The spectra in the paramagnetic state broaden below $T_{N}$ and take a rectangular shape, due to the internal field $H_{\rm int}$.}
\end{figure}
\subsection{NMR spectra below $T_{N}$}
Below $T_{\rm N}$, the $^{31}$P spectra measured at 75.5\,MHz were found to broaden abruptly. In order to precisely probe the intrinsic line shape, we remeasured the $^{31}$P spectra at a lower frequency of 9.4\,MHz. As shown in Fig.~\ref{belowTn}, the $^{31}$P line above $T_{N}$ remains narrow and immediately below $T_{N}$ it starts broadening indicating that the P site is
experiencing the static internal field in the ordered state through the hyperfine field between the P nuclei and the
ordered V$^{4+}$ moments. With decrease in temperature, the spectrum takes a nearly rectangular shape but the central peak still persists down to the lowest measured temperature. The relative intensity of the central peak with respect to the broad rectangular spectra decreases with decreasing temperatures. As discussed later, this central peak is found to be intrinsic to the sample.\cite{vonlanthen2002}

\begin{figure}
\includegraphics [width=3in]{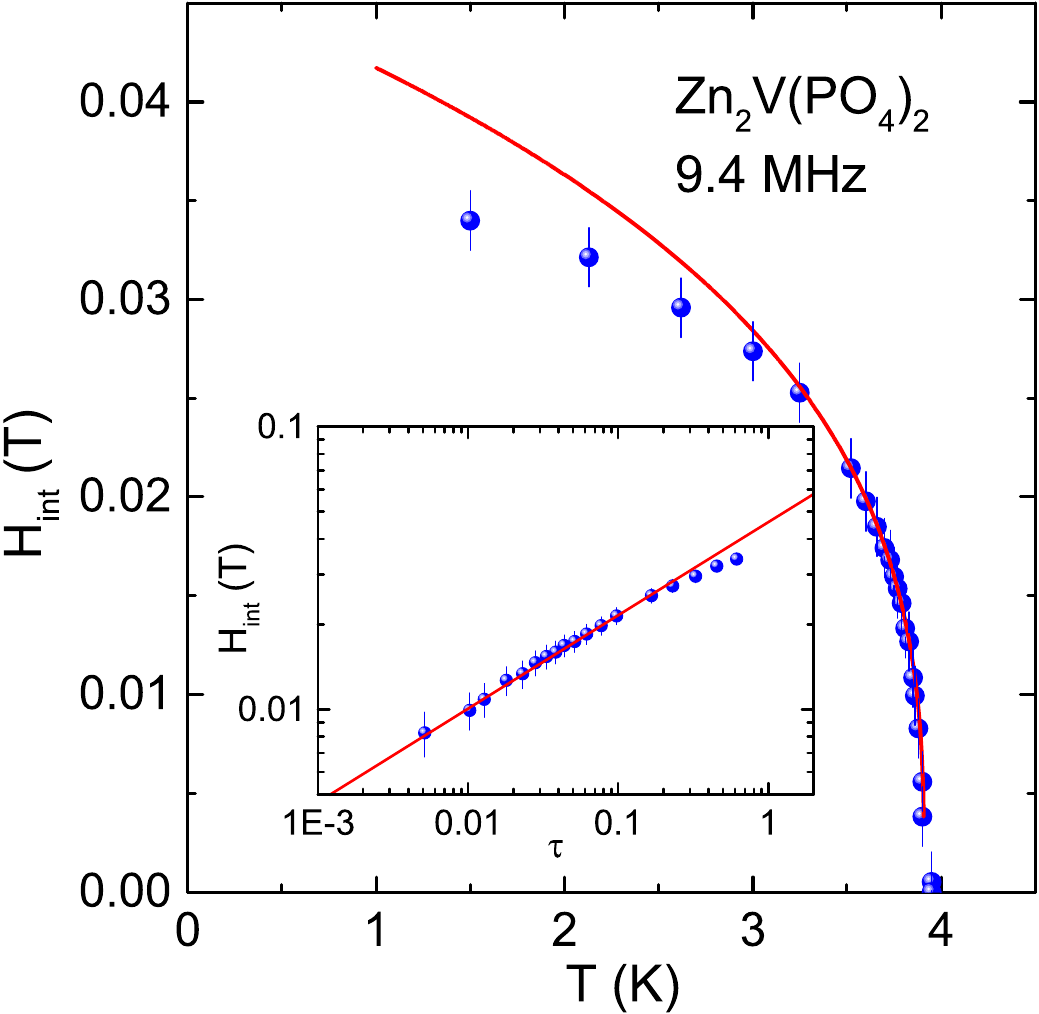}
\caption{\label{sublattice} (Color online) Temperature dependence of the internal field $H_{\rm int}$ obtained from NMR spectra
measured at 9.4\,MHz in the ordered state. $H_{\rm int}$ is proportional to the V$^{4+}$ sublattice magnetization. The solid line is the fit by Eq.~\eqref{ms} as described in the text. Inset: $H_{\rm int}$ vs. $\tau$ and the solid line is the simulation of $0.046 \times \tau^{0.33}$ taking $T_{\rm N} \simeq 3.90$~K.}
\end{figure}

The internal field $H_{\rm int}$, which is proportional to the V$^{4+}$ sublattice magnetization, was determined by taking the half width at the half maximum from the fit of the experimental spectra following the procedure adopted recently for BiMn$_2$PO$_6$ (Ref.~\onlinecite{nath2014}). The temperature dependence of $H_{\rm int}$ is plotted in Fig.~\ref{sublattice}. In order to extract the critical exponent ($\beta$) of the order parameter (sublattice magnetization), $H_{\rm int}(T)$ was fitted by the power law:
\begin{equation}
H_{\rm int}(T)=H_{0}\left(1-\frac{T}{T_{\rm N}}\right)^{\beta}.
\label{ms}
\end{equation}
One can notice that $H_{\rm int}$ decreases sharply on approaching $T_{\rm N}$. For a precise estimation of $\beta$, one needs more data points close to $T_{\rm N}$. We have estimated $\beta$ by fitting the data points as close as possible to $T_{\rm N}$ (i.e., in the critical region) as shown in Fig.~\ref{sublattice}. The maximum value of $\beta = 0.33 \pm 0.02$ with $H_0 \simeq 0.046(2)$~T and $T_{\rm N} \simeq 3.9(1)$\,K was obtained by fitting the data points in the $T$-range 3.7\,K to 3.95\,K close to $T_{\rm N}$. By increasing the number of fitting points toward low-$T$s, the $\beta$ value was found to decrease. In order to magnify the fit in the critical region, $H_{\rm int}$ is plotted against the reduced temperature $\tau = 1-\frac{T}{T_{\rm N}}$ in the inset of Fig.~\ref{sublattice}. The solid line is the fit by $0.046 \times \tau^{0.33}$ where $T_{\rm N}$ is taken to be 3.90~K. At low-$T$s, $H_{\rm int}$ develops the tendency of saturation and it saturates faster than expected from the mean-field theory [see the deviation of fits in Fig.~\ref{sublattice} at low-$T$s].

\subsection{Nuclear spin-lattice relaxation rate $1/T_{1}$}
\begin{figure}
\includegraphics [width=4in]{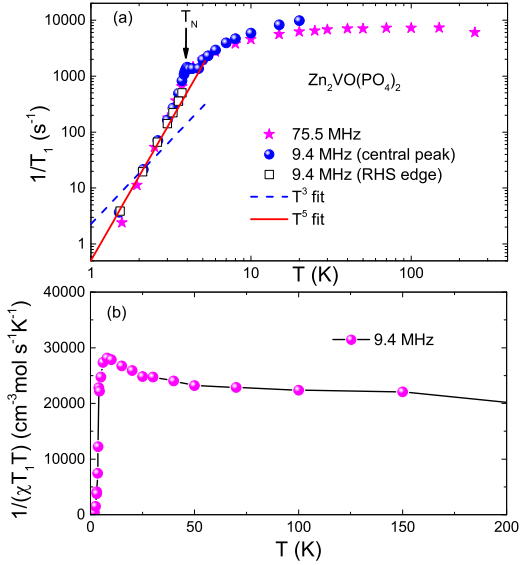}
\caption{\label{t1} (Color online) (a) Spin-lattice relaxation rate $1/T_{1}$ vs. temperature $T$ measured at 75.5 and 9.4\,MHz. Two data sets at 9.4\,MHz correspond to the measurements at both the central peak and RHS edge positions below $T_{\rm N}$ [see Fig.~\ref{belowTn}]. The solid and dashed lines represent $T^5$ and $T^3$ behaviors, respectively. (b) $1/(\chi T_{1}T)$ is
plotted as a function of $T$.}
\end{figure}
The $^{31}$P nuclear spin-lattice relaxation rate $1/T_{1}$ above $T_{\rm N}$ was measured at the field corresponding to the central peak position. For $T\leq T_{\rm N}$, the measurements were performed at both the central peak position as well as at the right-hand side (RHS) edge position (see Fig.~\ref{belowTn}). For an $I=1/2$ nucleus, the recovery of the longitudinal magnetization is expected to follow a single-exponential behavior. In Zn$_{2}$VO(PO$_{4}$)$_{2}$, the recovery of the longitudinal nuclear magnetization was indeed fitted well by the exponential function $1-\frac{M(t)}{M_{0}}=Ae^{-t/T_{1}}$,
where $M(t)$ is the nuclear magnetization at a time $t$ after the saturation pulse and $M_{0}$ is the equilibrium magnetization.
The temperature dependence of $1/T_{1}$ extracted from the fit is presented in Fig.~\ref{t1}(a).

The $1/T_{1}$ data measured at two different frequencies (75.5\,MHz and 9.4\,MHz) almost resemble each other at low temperatures.
At high temperatures ($T \gtrsim 10$~K), $1/T_{1}$ is temperature-independent. In the high temperature limit $T\gg J/k_{\rm B}$,
a temperature-independent $1/T_{1}$ behavior is typical due to random fluctuation of paramagnetic moments.\cite{moriya1956} With decrease in temperature, $1/T_{1}$ decreases slowly for $T<10$~K and then shows a weak anomaly around $T_{\rm N}\simeq 3.8$~K. This decrease is very similar to that observed previously in the cases of the antiferromagnetic square lattices Pb$_{2}$VO(PO$_{4}$)$_{2}$ (Ref.~\onlinecite{nath2009}), SrZnVO(PO$_4$)$_2$,(Ref.~\onlinecite{bossoni2011}), VOMoO$_{4}$ (Ref.~\onlinecite{carretta2002b}), and [Cu(HCO$_{2}$)$_{2}$.4D$_{2}$O], where the decrease of $1/T_{1}$ above $T_{\rm N}$ is explained by cancellation of the antiferromagnetic spin fluctuations at the probed nuclei.\cite{carretta2000}

Below the peak, $1/T_{1}$ again decreases smoothly towards zero. As shown in Fig.~\ref{t1}(a) no difference in $1/T_{1}$ below $T_{\rm N}$ was observed between the data measured at the central peak and RHS edge positions at 9.4\,MHz.

\section{Ti-doped Zn$_2$VO(PO$_4$)$_2$}
\begin{figure}
\includegraphics{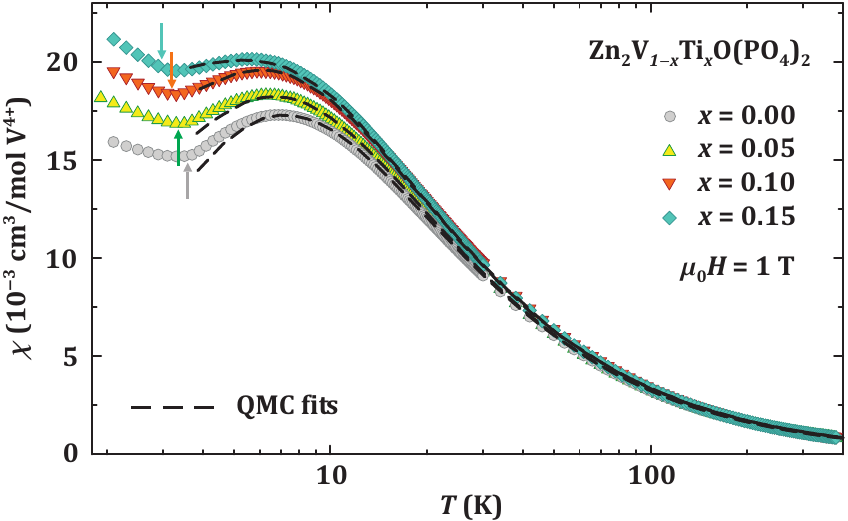}
\caption{\label{fig:chi-doped} (Color online) Magnetic susceptibility of Ti-doped Zn$_2$VO(PO$_4$)$_2$ measured at $\mu_0H=1$\,T. The dashed lines are QMC fits with the diluted square-lattice model, as described in the text. The arrows mark N\'eel temperatures $T_N$ that systematically decrease upon doping (see also Fig.~\ref{fig:heat-doped}). }
\end{figure}

As mentioned in Sec.~\ref{sec:methods}, all Ti-doped samples revealed tetragonal symmetry, similar to the parent compound. The sample with $x=0.15$ contained trace amounts of impurity phases, so its doping level may be slightly below 15\%, but this minor deviation had no visible effect on the results.

\begin{figure}
\includegraphics{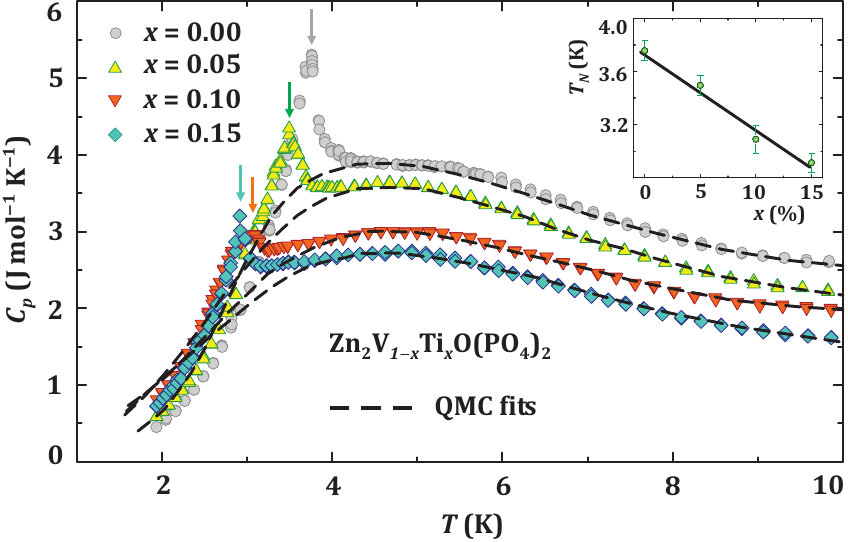}
\caption{\label{fig:heat-doped} (Color online) Specific heat of Ti-doped Zn$_2$VO(PO$_4)_2$ measured in zero magnetic field. The dashed lines are QMC fits, and the arrows mark N\'eel temperatures $T_N$ depicted in the inset as a function of the doping level $x$. The solid line in the inset is the tentative linear fit $T_N=(1-Cx)T_N(0)$ with $C=1.5$}
\end{figure}

Magnetic susceptibility of doped samples normalized to one mole of V$^{4+}$ spins is shown in Fig.~\ref{fig:chi-doped}. The susceptibility maximum is systematically shifted to higher values of $\chi$ and to lower temperatures. For the sake of better presentation, we use a different scaling for the specific heat and normalize the data to one mole of the compound. Fig.~\ref{fig:heat-doped} presents the systematic reduction in the specific heat maximum around 4.5\,K following the reduced amount of the magnetic V$^{4+}$ ions. The position of the maximum is roughly unchanged up to $x=0.15$.

Magnetic order persists in all Ti-doped samples. The magnetic transition is seen by a change in the slope of $\chi(T)$ (Fig.~\ref{fig:chi-doped}). The precise value of $T_N$ is better tracked by the $\lambda$-type anomaly in the specific heat (Fig.~\ref{fig:heat-doped}). The N\'eel temperature determined with the 0.05\,K uncertainty from the maximum of the transition anomaly, displays a systematic reduction from 3.8\,K in the parent compound to 2.9\,K at $x=0.15$. This corresponds to the slope of $-dT_N(x)/dx=CT_N(0)$ with $C=1.5(2)$, which is reminiscent of $C\simeq 2$ in Li$_2$VOSiO$_4$\cite{papinutto2005} and well below $C\simeq 2.7$ or $C\simeq 3.5$ for La$_2$CuO$_4$ doped with Mg and Zn, respectively.\cite{carretta2011}

\begin{table}
\caption{\label{tab:fits}
Parameters obtained from fitting the susceptibility and specific heat data for Zn$_2$V$_{1-x}$Ti$_x$O(PO$_4)_2$ with QMC results for the ideal ($x=0$) and diluted ($x>0$) square-lattice models. $g$ stands for the $g$-factor, $\chi_0$ is the temperature-independent contribution to the susceptibility (in~$10^{-5}$\,emu/mol), and $J$ is the exchange coupling (in~K).
}
\begin{ruledtabular}
\begin{tabular}{c@{\hspace{1em}}ccc@{\hspace{1em}}c}
  $x$ & \multicolumn{3}{c}{Susceptibility} & Specific heat \\
      &  $g$  & $\chi_0$ &      $J$        &     $J$       \\\hline
 0.00 & 1.95  &  $-5$    &      7.7        &     7.8       \\
 0.05 & 1.95  &  $-6$    &      7.7        &     7.8       \\
 0.10 & 1.97  &  $-4$    &      7.8        &     7.9       \\
 0.15 & 1.97  &  $-6$    &      8.1        &     8.0       \\
\end{tabular}
\end{ruledtabular}
\end{table}

N\'eel temperature of an antiferromagnet depends on its exchange couplings. Therefore, for a proper interpretation of $T_N(x)$ and its slope, one has to evaluate the change in $J$ upon doping. To this end, we fitted magnetic susceptibility and specific heat of Ti-doped samples in the same manner as we did in Sec.~\ref{sec:thermo} for the parent compound. Model curves were obtained by QMC simulations for the diluted square lattice of spins-$\frac12$. Fitted parameters are listed in Table~\ref{tab:fits} and show a good match between the susceptibility and specific heat data. The error bar for the values of $J$ is somewhat difficult to define, because statistical errors largely depend on the temperature range of the fitting. However, even with a very optimistic error bar of 0.1\,K for the susceptibility fits above 7\,K, the change in $J$ between $x=0$ and $x=0.15$ is only marginal. Moreover, $T_N$ depends on $\ln J$,\cite{yasuda2005} so the 4\% change in the $J$ value should have negligible effect on the $T_N$. Its reduction is, therefore, solely due to the dilution, and the slope of $T_N(x)$ reflects the dilution effect on the spin-$\frac12$ AFM square lattice in Zn$_2$VO(PO$_4)_2$.

\section{Discussion}
\subsection{Static Properties}
The exchange couplings extracted from $\chi(T)$, $I_{\rm ESR}(T)$, and $K_{\rm iso}(T)$ data for Zn$_{2}$VO(PO$_{4}$)$_{2}$
are consistent with the values reported before from the $\chi(T)$
analysis\cite{kini2006} and neutron scattering experiments within the error bar.\cite{yusuf2010} According to the $J_{2}/J\simeq 0.03$ ratio, Zn$_{2}$VO(PO$_{4}$)$_{2}$ features the N\'eel antiferromagnetic ground state with antiparallel spins on nearest neighbors in the $ab$ plane (Fig.~\ref{fig:structure}, right).\cite{yusuf2010} In the crystal structure, squares are formed via V--O--P--O--V superexchange interaction paths. In contrast to Pb$_{2}$VO(PO$_{4}$)$_{2}$ where each P atom is coupled to four V$^{4+}$ spins,\cite{nath2009} in Zn$_{2}$VO(PO$_{4}$)$_{2}$ each P atom is coupled to two V$^{4+}$ spins only (Fig.~\ref{fig:structure}, right).

The total hyperfine coupling constant at the P site is the sum of transferred hyperfine ($A_{\rm trans}$) and dipolar ($A_{\rm dip}$) couplings produced by V$^{4+}$ spins, i.e., $A_{\rm hf}=z^{'}A_{\rm trans}+A_{\rm dip}$, where $z^{'}=2$ is the number of nearest-neighbor V$^{4+}$ spins of the P site. The anisotropic dipolar couplings were calculated using lattice sums to be $A_{\rm a} = 210~$Oe/$\mu_{\rm B}$, $A_{\rm b} = 210$~Oe/$\mu_{\rm B}$, and $A_{c} = -420$~Oe/$\mu_{\rm B}$ along the $a$-, $b$-, and $c$-directions, respectively. Clearly, the value of dipolar coupling is almost negligible compared to the total hyperfine coupling [$A_{\rm hf}^{\rm iso} = (9221 \pm 100)$ Oe/$\mu_{\rm B}$] suggesting that the dominant contribution to the total hyperfine coupling is due to the transferred hyperfine coupling at the P site. The magnitude of this coupling depends on the relative orientation and the extent of overlap between the V($3d$), P($2p$), and O($2s$) orbitals.

The internal field at the P site will be canceled out if the P ion is located at a symmetric position with respect to two nearest neighbor V$^{4+}$ up and down spins. However, the observation of a small remnant internal field at the P sites in the AFM ordered state indicates that the P sites are not located at the perfect symmetric position and there is a small displacement of the P sites from the perfect symmetric position. This is also consistent with the crystal structure where the P is sitting slightly above or below the line joining the neighboring up and down spins (see the right panel of Fig.~\ref{fig:structure}).
The $^{31}$P line in the magnetically ordered state takes a typical rectangular shape, reflecting that the magnetic ordering is commensurate in nature. If the magnetic structure were incommensurate with the lattice, the internal field would be distributed and the spectrum would not exhibit the rectangular shape seen in Fig.~\ref{belowTn}.
Our spectra are, therefore, consistent with the collinear magnetic order determined from the neutron diffraction experiments.\cite{yusuf2010}

The central line does not disappear from the experimental spectra completely even at the lowest measured temperature. NMR experiments on many other compounds, especially on powder samples, are reported to show similar coexistence of the high-$T$ and low-$T$ phases, e.g., in BaCuP$_2$O$_7$ (Ref.~\onlinecite{nath2005}), (Li,Na)VGe$_2$O$_6$ (Refs.~\onlinecite{gavilano2000,vonlanthen2002,pedrini2004}), (Ca$_4$Al$_2$O$_6$)Fe$_2$(As$_{1-x}$P$_x$)$_2$ (Ref.~\onlinecite{kinouchi2013}), BiMn$_2$PO$_6$ (Ref.~\onlinecite{nath2014}), and LiGaCr$_4$O$_8$ (Ref.~\onlinecite{tanaka2014}). The origin of this central line remains an open question.
One could argue that the coexistence of two phases is due to a spread of the transition temperatures within the polycrystalline sample, but in such a case it is quite unlikely to observe a distinct peak in the temperature dependence of $1/T_1$, as seen in Fig.~\ref{t1}.

One possible origin of the central line is the impurity phases. In order to check that, we measured $1/T_{1}$ below $T_{\rm N}$ at the positions corresponding to the central peak and the RHS edge of the spectra. It is to be noted that for any phosphorus containing impurity phase, the corresponding $T_1$ is expected to be different from the intrinsic $T_1$ of the sample. Moreover, if the central peak is a superposition of intrinsic and extrinsic contributions, one would observe a double exponential behaviour of the longitudinal recovery curves. However our recovery curves at both positions follow single exponential behaviour and the magnitude and the temperature dependence of $1/T_{1}$ at both positions are the same, which clearly suggests that the central peak is an intrinsic feature of the sample and completely rules out the contribution of impurity phases.
As discussed earlier, the P site in the ordered state experiences a finite internal field due to a slightly asymmetric position with respect to the neighboring up- and down- spins. On the contrary, a perfectly symmetric position of P should results in a single narrow spectral line at the zero-shift position. Hence it appears that the central peak may be originating from some P sites which are located close to the perfect symmetric position.
Another possible origin of the central line could also be the presence of crystal defects or local dislocations in the polycrystalline sample. NMR on a high-quality single crystal can probably resolve this issue.

The temperature dependence of $H_{\rm int}$ in the critical region provides the critical exponent $\beta$ reflecting the universality class of the spin system. The $\beta$ values expected for different spin- and lattice-dimensionalities are
listed in Table~II of Ref.~\onlinecite{nath2009}. The value of $\beta$ obtained from the experiment is $\approx 0.33$, which would be consistent with any of the 3D spin models (Heisenberg, Ising, or XY). Given the direction of spins along the $c$-axis in the magnetically ordered state,\cite{yusuf2010} the 3D Ising case looks plausible. On the other hand, the 3D behavior in the vicinity of $T_N$ should not be confused with the 2D-like behavior above $T_N$, where the data are well described by the 2D model and 2D spin correlations manifest themselves in neutron scattering.\cite{yusuf2010} However, the critical exponent for the 2D Ising model\cite{Collins1989,*OzekiR149} $\beta=\frac18$ would not be consistent with the experiment.

Given the fact that below $T_N$ spins are aligned with the $c$ direction,\cite{yusuf2010} we may expect a weak Ising anisotropy, but it is impossible to quantify this putative anisotropy using the data at hand. Interestingly, the critical behavior of Zn$_2$VO(PO$_4)_2$ deviates from that of other square-lattice V$^{4+}$ antiferromagnets, where $\beta\simeq 0.25$ (2D XY universality class) was systematically observed in Li$_2$VOSiO$_4$ and Li$_2$VOGeO$_4$ (Refs.~\onlinecite{melzi2000,*melzi2001,carretta2002}), Pb$_2$VO(PO$_4)_2$ (Refs.~\onlinecite{nath2009,forster2013,carretta2009}), SrZnVO(PO$_4)_2$ (Refs.~\onlinecite{bossoni2011,forster2014}), and other compounds.\cite{carretta2009} The origin of this difference should be addressed in future studies.

\subsection{Dynamic Properties}
As shown in Fig.~\ref{t1}(b), $1/(\chi T_{1}T)$ above $\sim 10$\,K is $T$-independent and increases slowly below 10\,K where the
system begins to show antiferromagnetic short-range order. The general expression for $\frac{1}{T_{1}T}$ in terms of the dynamic susceptibility $\chi_{M}(\vec{q},\omega_{0})$ is\cite{moriya1963,mahajan1998}
\begin{equation}
\frac{1}{T_{1}T} = \frac{2\gamma_{N}^{2}k_{B}}{N_{\rm A}^{2}}
\sum\limits_{\vec{q}}\mid A(\vec{q})\mid
^{2}\frac{\chi^{''}_{M}(\vec{q},\omega_{0})}{\omega_{0}},
\label{t1form}
\end{equation}
where the sum is over wave vectors $\vec{q}$ within the first Brillouin zone, $A(\vec{q})$ is the form factor of the hyperfine interactions as a function of $\vec{q}$, and $\chi^{''}_{M}(\vec{q},\omega _{0})$ is the imaginary part of the
dynamic susceptibility at the nuclear Larmor frequency $\omega _{0}$. For $q=0$ and $\omega_{0}=0$, the real component of $\chi_{M}^{'}(\vec{q},\omega _{0})$ corresponds to the uniform static susceptibility $\chi$. Thus the temperature-independent $1/(\chi T_{1}T)$ above 10\,K in Fig.~\ref{t1}(b) demonstrates the dominant contribution of $\chi$ to $1/T_{1}T$. On the other hand, a slight increase in $1/(\chi T_{1}T)$ below 10\,K indicates the growth of antiferromagnetic correlations with decreasing $T$.

The symmetric location of phosphorous between the two V$^{4+}$ spins implies that N\'{e}el-type AFM spin fluctuations [$\vec{q}=(\pm \pi/a, \pm \pi/b)$] from neighboring spins should be largely filtered out (${|A(\vec{q})|}^2=0$) because the P nuclei interact with V$^{4+}$ spins having opposite directions (Fig.~\ref{fig:structure}, right). When the coupling to the two V$^{4+}$ spins is equivalent, the AFM fluctuations do not contribute to $1/(\chi T_{1}T)$. The residual enhancement of $1/(\chi T_{1}T)$ below 10\,K reflects the asymmetry of the hyperfine couplings. This asymmetry is consistent with the crystal structure of Zn$_2$VO(PO$_4)_2$, where the P atoms are located on mirror planes running perpendicular to the ($\av+\bv$) or ($\av-\bv$) crystallographic directions. The tensor of hyperfine couplings may change its orientation upon the reflection in the mirror plane, thus leading to non-equivalent interactions between P and the up- and down-spins on the neighboring V$^{4+}$ ions.

In the AFM ordered state, $1/T_{1}$ is mainly driven by scattering of magnons, leading to
a power-law temperature dependence.\cite{beeman1968,belesi2006} For $T \gg \Delta/k_{\rm B}$,
where $\Delta/k_{\rm B}$ is the gap in the spin-wave spectrum, $1/T_{1}$ either follows a $T^{3}$ behavior due to a two-magnon
Raman process or a $T^{5}$ behavior due to a three-magnon process, while for $T \ll \Delta/k_{\rm B}$, it follows an activated behavior $1/T_{1} \propto T^{2}\exp(-\Delta/k_{\rm B}T)$. As seen from Fig.~\ref{t1}(a), our $^{31}$P $1/T_{1}$ data in the lowest temperature region (1.5\,K $\leq T \leq$ 3.25\,K) follow the $T^{5}$ behavior rather than the $T^{3}$ behavior suggesting that the relaxation is mainly governed by the three-magnon process. The lack of activated behavior down to 1.5\,K indicates that the upper limit of $\Delta/k_{\rm B}$ is 1.5\,K.

At sufficiently high temperatures, $1/T_{1}$ due to local moments is $T$-independent and can be expressed within the Gaussian approximation of the auto-correlation function of the electronic spin as:\cite{moriya1956}
\begin{equation}
\left(\frac{1}{T_1}\right)_{T\rightarrow\infty} =
\frac{(\gamma_{N} g\mu_{\rm B})^{2}\sqrt{2\pi}z^\prime S(S+1)}{3\,\omega_{ex}}
{\left(\frac{A_{hf}}{z'}\right)^{2}},
\label{t1inf}
\end{equation}
where $\omega_{ex}=\left(|J_{\rm max}|k_{\rm B}/\hbar\right)\sqrt{2zS(S+1)/3}$ is the Heisenberg exchange frequency, $z$ is the number of nearest-neighbor spins of each V$^{4+}$ ion, and $z^\prime$ is the number of nearest-neighbor V$^{4+}$ spins for a given P site. The $z^\prime$ in the numerator takes into account the number of
nearest-neighbor V$^{4+}$ spins responsible for producing fluctuations at the P site. Using the relevant parameters, $A_{\rm hf} \simeq 9221$\,Oe/$\mu_{\rm B}$, $\gamma_N = 1.08 \times 10^8\,{\rm rad}$~s$^{-1}$\,T$^{-1}$, $z=4$, $z^\prime=2$, $g=2$, $S=\frac12$, and the high-temperature (150\,K) relaxation rate of
$\left(\frac{1}{T_1}\right)_{T\rightarrow\infty}\simeq 7270.6$~s$^{-1}$ for the P site in Eq.~\eqref{t1inf}, the magnitude of the exchange coupling is calculated to be $J\simeq 9$\,K in good agreement with $J\simeq 7.7$\,K determined from the thermodynamic data (Sec.~\ref{sec:thermo}).
\begin{figure}
\includegraphics [width=3in]{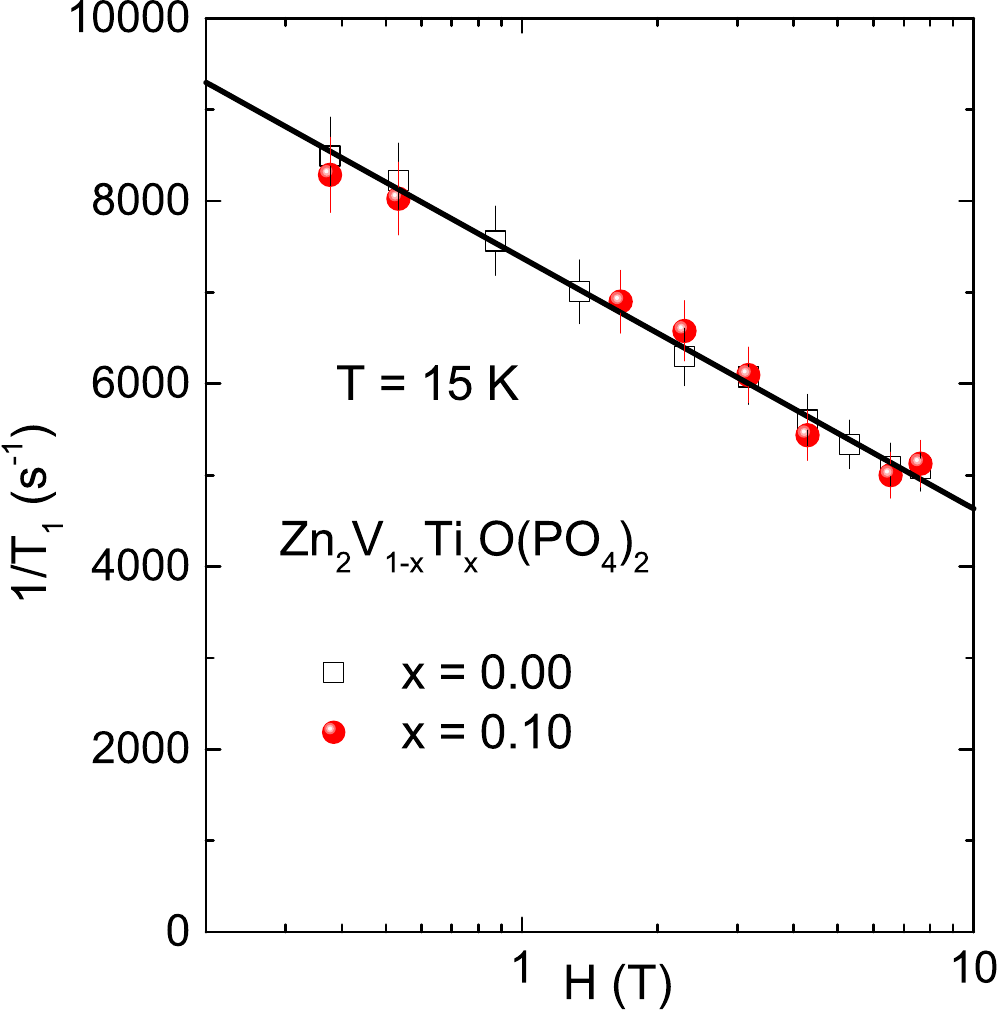}
\caption{\label{T1vsH} (Color online) $1/T_1$ vs. $H$ (in log scale) measured at $T=15$~K for Zn$_2$V$_{1-x}$Ti$_x$O(PO$_4$)$_2$ ($x$ = 0 and 0.10) samples. The solid line is the fit by $1/T_1=a+b\log(1/H)$.}
\end{figure}

One can see in Fig.~\ref{t1}(a) that for $T>10$\,K a slight increase in $1/T_1$ was observed at 9.4\,MHz compared to the data measured at 75.5\,MHz. In order to check whether this difference is due to the effect of spin diffusion, we measured $1/T_1$ at different applied fields at $T=15$\,K. Long-wavelength ($q\sim0$) spin fluctuations in a Heisenberg magnet show diffusive
dynamics. In 1D compounds, such spin diffusion results in a $1/\sqrt{H}$ magnetic field dependence of $1/T_1$, which has been
observed in (CH$_3$)$_4$NMnCl$_3$, CuCl$_2$.2NC$_5$H$_5$, and Sr$_2$CuO$_3$.\cite{hone1974,ajiro1978,takigawa1996}
On the other hand, in 2D materials $1/T_1$ varies as $\log(1/H)$.\cite{furukawa1996, ajiro1978}

In Fig~\ref{T1vsH}, $1/T_1$ is plotted against $H$ (in log scale) measured at $T=15$~K for Zn$_2$V$_{1-x}$Ti$_x$O(PO$_4$)$_2$ ($x$ = 0 and 0.10) samples. Both the data sets resemble with each other and show the same field dependency. They can be fitted by the form $1/T_1=a+b\log(1/H)$ where $a$ and $b$ are constants. The linearity of the $1/T_1$ vs. $\log(H)$ dependence is indicative of the 2D nature of both the parent and 10\% Ti$^{4+}$ doped samples above $T_N$.

\subsection{Effect of doping}
Zn$_2$VO(PO$_4)_2$ reveals a very clean case of a diluted antiferromagnet. We have shown that the change in the nearest-neighbor coupling $J$ is marginal (Table~\ref{tab:fits}), and the frustration by second-neighbor couplings $J_2$ is negligible in the parent compound. However, the N\'eel temperature of Zn$_2$VO(PO$_4)_2$ drops much slower than expected for the diluted AFM square lattice of spins-$\frac12$. In a diluted system, the $T_N$ can be written as follows:\cite{chen2000,papinutto2005}
\begin{equation}
 k_BT_N(x)=J_{\perp}(1-x)^2\xi(x,T_N)^2\left(\frac{M(x)}{M(0)}\right)^2,
\end{equation}
where $J_{\perp}(1-x)^2$ reflects the reduction in the interlayer coupling (the probability to find two coupled spins in the adjacent layers), $\xi(x,T_N)$ is the in-plane correlation length, and $M(x)$ is the staggered magnetization at a given value of $x$. All these factors taken together should yield the slope $C\simeq 3.2$ (Ref.~\onlinecite{chernyshev2002}) for the linear dependence of $T_N(x)$ and spin-$\frac12$. Experimentally, Carretta~\textit{et~al.}\cite{carretta2011} report $C\simeq 2.7$ and 3.5 for Mg- and Zn-doped La$_2$CuO$_4$, respectively.\cite{delannoy2009} This finding can be rationalized by assuming that Zn atoms introduce additional frustration, whereas Mg atoms do not.\cite{liu2009,carretta2011}

In Ti-doped Li$_2$VOSiO$_4$, the slope of $T_N(x)$ is $C\simeq 2$, only. Papinutto~\textit{et~al.}\cite{papinutto2005} proposed that this slope is solely due to the first term $J_{\perp}(1-x)^2$, while $M(x)$ is only weakly influenced by doping because the effect of dilution is countered by the change in the frustration ratio. This explanation looks plausible for Li$_2$VOSiO$_4$ indeed, because the physics of this compound is determined by the competing nearest-neighbor and second-neighbor couplings on the square lattice.\cite{rosner2002,*rosner2003,melzi2000,*melzi2001} Ti-doped Zn$_2$VO(PO$_4)_2$ reveals an even lower $C\simeq 1.5$, and in this compound frustration is clearly inactive. We have shown that the frustration is vanishingly small ($J_2/J\simeq 0.03$) in the pristine Zn$_2$VO(PO$_4)_2$, while its increase (if any) will have an opposite effect on the system and increase $C$ above 3.2 instead of decreasing it to the experimental $C\simeq 1.5$ value.

The different doping behavior of Zn$_2$VO(PO$_4)_2$ and Li$_2$VOSiO$_4$ on one hand and La$_2$CuO$_4$ on the other can be ascribed to a different magnitude of their interlayer couplings. While Zn$_2$VO(PO$_4)_2$ shows signatures of the 2D physics above $T_N$, the N\'eel temperature of this compound is quite high, $T_N/J\simeq 0.5$, hence $|J_{\perp}|/J\simeq 10^{-1}$. In Li$_2$VOSiO$_4$, the lower N\'eel temperature of $T_N/J\simeq 0.32$ corresponds to an order-of-magnitude weaker interlayer coupling $|J_{\perp}|/J\simeq 10^{-2}$,\cite{melzi2000,*melzi2001,yasuda2005} which is still much stronger than in La$_2$CuO$_4$ with its $T_N/J\simeq 0.21$ and $J_{\perp}/J\ll 10^{-3}$. 

Magnetic anisotropy could be another reason for the different evolution of $T_N$ upon doping, but its effect is difficult to quantify. In La$_2$CuO$_4$, Dzyaloshinsky-Moriya (DM) terms, the leading component of the anisotropy in spin-$\frac12$ magnets, are about 1.5\% of $J$.\cite{birgeneau1999} Crystallographic symmetries of both Li$_2$VOSiO$_4$ and Zn$_2$VO(PO$_4)_2$ allow for non-zero DM couplings as well, but their magnitude is presently unknown. Regarding Zn$_2$VO(PO$_4)_2$, our NMR data provide an upper threshold of about 1.5\,K for the anisotropy gap. This value is, however, nearly 20\% of $J$ and exceeds typical DM anisotropies in V$^{4+}$ oxides.\cite{lumsden2001,*ivanshin2003}

The variable interlayer coupling is a plausible reason for the different doping evolution of $T_N$ in square-lattice antiferromagnets. In La$_2$CuO$_4$, the long-range order emerges only at low temperatures where the in-plane correlation length is about 100 lattice spacings,\cite{birgeneau1999,greven1995} and the magnetic order is vulnerable to the dilution and disorder. In Li$_2$VOSiO$_4$ and especially in Zn$_2$VO(PO$_4)_2$, the in-plane correlation length at $T_N$ is on the order of several lattice spacings, and interlayer couplings have larger influence on the long-range ordering, thus reducing the slope of $T_N(x)$ compared to the ideal 2D case where $J_{\perp}\ll J$. Therefore, the doping scenario of Zn$_2$VO(PO$_4)_2$ may be of 3D type and will require one to view this compound as a spatially anisotropic 3D antiferromagnet, even though the physics above $T_N$ is 2D-like.\cite{kini2006,yusuf2010}

Finally, we note that our data do not support the \textit{ab initio} predictions by Kanungo~\textit{et~al.}\cite{kanungo2013} regarding the 1D physics of doped Zn$_2$VO(PO$_4)_2$. While probably correct for the ordered monoclinic structure at the 25\% doping level, their results do not apply to our case, where magnetic V$^{4+}$ and non-magnetic Ti$^{4+}$ ions are randomly distributed in the structure, and the overall tetragonal symmetry is retained.

\section{Summary and conclusions}
Zn$_2$VO(PO$_4)_2$ is an antiferromagnetic compound with the in-plane coupling of $J\simeq 7.7$\,K, negligible in-plane frustration, and long-range magnetic order below $T_N\simeq 3.75$\,K. Thermodynamic properties above $T_N$ are well described by the Heisenberg model on the AFM square lattice. NMR results confirm the commensurate nature of the magnetic order. The spin-lattice relaxation rate $1/T_1$ below $T_{\rm N}$ follows the $T^{5}$ behavior reflecting that the relaxation is governed by the three-magnon process. $1/T_1$ at 15\,K varies as $\log(1/H)$ and supports the presence of strong 2D spatial anisotropy in both the parent and 10\% Ti$^{4+}$ doped compounds above $T_N$. On the other hand, the critical exponent for the sublattice magnetization is consistent with any of the 3D universality classes and may reflect the sizeable interlayer exchange in Zn$_2$VO(PO$_4)_2$. Ti$^{4+}$ doping with up to 15\% of Ti$^{4+}$ leads to a uniform dilution of the spin lattice and only a marginal change in the in-plane exchange coupling. $T_N$ goes down in a linear manner, but its slope is well below theoretical expectations for the diluted Heisenberg antiferromagnet on the square lattice of spins-$\frac12$ and may indiciate the importance of the interlayer exchange.

\begin{acknowledgments}
AY, NA, and RCN would like to acknowledge DST India for financial support. AT was funded by the Mobilitas program of the ESF (grant No. MTT77) and by the IUT23-3 grant of the Estonian Research Agency. Work at the Ames Laboratory was supported by the Department of Energy-Basic Energy Sciences under Contact No. DE-AC02-07CH11358. Fig.~\ref{fig:structure} was prepared using the \texttt{VESTA} software.\cite{vesta}
\end{acknowledgments}

\end{document}